\documentclass[letterpaper,12pt]{article}
\usepackage{epsfig,colortbl,multicol}
\usepackage{graphicx}
\usepackage{natbib}
\newcommand{\teff}{$T_{\rm eff}$}
\newcommand{\cp}{\citep}
\newcommand{\ct}{\citet}
\newcommand{\apj}{ApJ}
\newcommand{\aj}{AJ}
\newcommand{\apjl}{ApJ}
\newcommand{\aap}{A\&A}
\newcommand{\nat}{Nature}
\newcommand{\apjs}{ApJS}
\newcommand{\mnras}{MNRAS}
\newcommand{\prb}{Phys Rev B}
\newcommand{\planss}{P\&SS}

\newcommand{\jgr}{JGR}
\newcommand{\apss}{Astrophys and Space Sci}

\begin{document}
\begin{center}
\Large
\textbf{{Planetary Formation and Evolution Revealed with a Saturn Entry Probe: The Importance of Noble Gases}}\\
\large
\end{center}

\begin{multicols}{2}
\noindent{Jonathan J. Fortney, UC Santa Cruz}\\
\texttt{jfortney@ucolick.org}\\
Kevin Zahnle, NASA Ames\\ 
Isabelle Baraffe, CRAL, Lyon\\
Adam Burrows, Princeton University\\
Sarah E. Dodson-Robinson, Caltech\\
Gilles Chabrier, CRAL, Lyon\\
Tristan Guillot, Nice Observatory\\
Ravit Helled, UCLA\\
Franck Hersant, Bordeaux Observatory\\
William B. Hubbard, University of Arizona\\
Jack J. Lissauer, NASA Ames\\
Mark S. Marley, NASA Ames\\

\end{multicols}

The determination of Saturn's atmospheric noble gas abundances are critical to understanding the formation and evolution of Saturn, and giant planets in general.  These measurements can only be performed with an entry probe.  A Saturn probe will address whether enhancement in heavy noble gases, as was found in Jupiter, are a general feature of giant planets, and their ratios will be a powerful constraint on how they form.  The helium abundance will show the extent to which helium has phase separated from hydrogen in the planet's deep interior.  Jupiter's striking neon depletion may also be tied to its helium depletion, and must be confirmed or refuted in Saturn.  Together with Jupiter's measured atmospheric helium abundance, a consistent evolutionary theory for both planets, including ``helium rain'' will be possible.  We will then be able to calibrate the theory of the evolution of all giant planets, including exoplanets.  In addition, high pressure H/He mixtures under giant planet conditions are an important area of condensed matter physics that are beyond the realm of experiment.

\section{Giant Planet Evolution}
\subsection{The Standard Giant Planet Picture: Jupiter}
There is a standard theory of giant planet interior structure that goes back to the pioneering work of Bill Hubbard in the late 1960s \cp{Hubbard68}.  \ct{Low66} found that Jupiter emits more mid-infrared radiation than it receives from the Sun, which directly implies that Jupiter possesses its own interior energy source.  \ct{Hubbard68} showed that Jupiter's intrinsic flux could not be carried throughout its interior by conduction or radiation, implying that convection dominated the energy transport in the planet's interior.  Efficient convection implied that Jupiter's interior temperature gradient is adiabatic, and that Jupiter is mostly composed of \emph{warm, fluid} hydrogen, not cold solid hydrogen, which was an open question at the time.

In the mid 1970s it was clearly deduced from Jupiter and Saturn's radii and gravity fields that these planets have heavy element cores \cp[e.g.][]{Podolak74}.  The first thermal evolution calculations of these planets were performed by a number of authors at around this time \cp{Graboske75,Bodenheimer76,Hubbard77}.  Models for Jupiter, starting from an initially hot state post-formation, with a H/He envelope that was assumed homogeneous, adiabatic, and well-mixed, reached Jupiter's known \teff\ of 124 K in $\sim$4.5 Gyr.  This helped to form the paradigm of the adiabatic, fully convective giant planet.

\subsection{Saturn: A More Complex Story}
These same kinds of evolution models could also be easily applied to Saturn.  However, these calculations failed badly to reproduce Saturn, as shown in Figure 1.  A Saturnian cooling age of 2-2.5 Gyr was found, implying that Saturn today (\teff=95 K) is much too hot, by a factor of 50\% in luminosity \cp{Pollack77,SS77a}.  This was the first, and still most important crack in the standard theory.

At around this same time Stevenson and Salpeter were examining in some detail the physics of H/He mixtures, and how phase separation between H and He (or ``demixing'') might effect the evolution of giant planets \cp{Salpeter73,Stevenson75,SS77b,SS77a}.  They found that a ``rain'' of helium was likely within Saturn, and perhaps Jupiter, and that the differentiation (a conversion of gravitational potential energy to thermal energy) could prolong Saturn's evolution, keeping it warmer, longer.  As the helium droplets separate out and rain from megabar pressures, eventually redissolving at higher pressures and temperatures, $Y$ is enhanced in the very deep interior.  Helium is also lost from the visible atmosphere because the entire planet above the rain region is convective and well mixed \cp{SS77a}.
\begin{figure}[hb!]
\begin{center}
\includegraphics[scale=0.85]{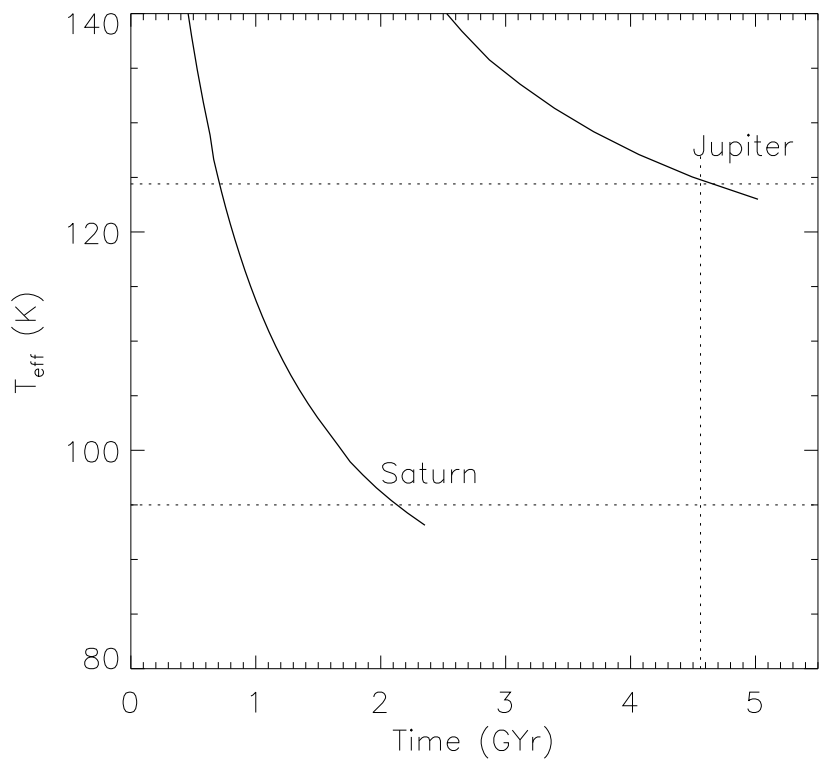}
\end{center}
\caption{\small Fully adiabatic, homogeneous H/He envelope models of the thermal evolution of Jupiter and Saturn, after \ct{FH03}.  These models include the energy input of the Sun with time.  For Saturn, the real planet (age 4.55 Gyr) has a much higher \teff\ than the model, indicating the model is missing important physics.}\label{js}
\end{figure}
%
%
\section{Giant Planet Formation}
The \emph{Galileo} mission made two deeply surprising discoveries.  One is that Callisto appears to have stopped at the brink of fully differentiating.  This is being addressed by Jupiter orbiters.  The other major surprise, from the \emph{Galileo Entry Probe}, is that the heavier noble gases Ar, Kr, and Xe appear to be significantly more abundant in the Jovian atmosphere than in the Sun, at enhancements generally comparable to what was seen for the chemically active volatiles N, C, and S.  It had been generally expected that Ar, Kr, and Xe would be present in solar abundances, as all were expected to accrete with hydrogen during the gravitational capture of nebular gases.  
\begin{figure}[hb!]
\begin{center}
\includegraphics[scale=1.2]{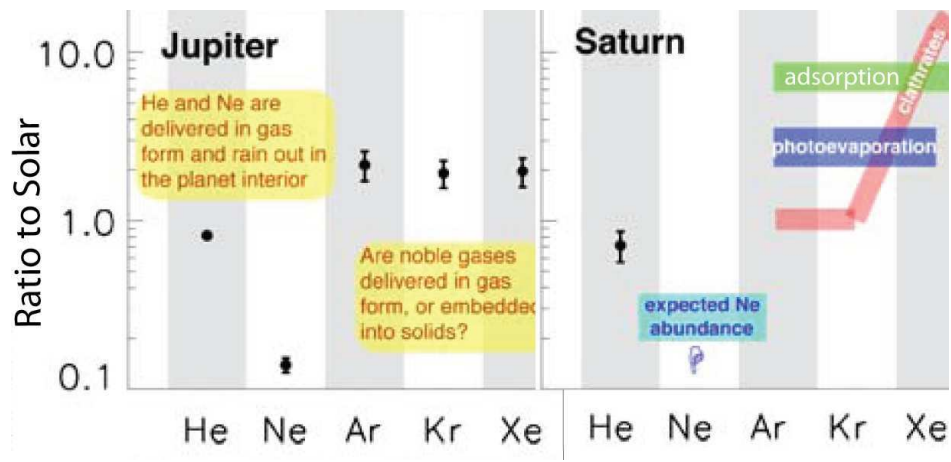}
\end{center}
\caption{\small Elemental abundances measured in the tropospheres of Jupiter (top) and Saturn (bottom) in units of their abundances in the protosolar nebula. The elemental abundances for Jupiter are derived from the in situ measurements of the Galileo probe. The (unpublished) He abundance for Saturn is from spectroscopic determinations from Cassini.  A Saturn probe will distinguish between different formation scenarios whose predictions are shown as green, blue, and pink curves, respectively.  Adapted from \ct{Marty09}}\label{kronos}
\end{figure}

Enhanced abundances of Ar, Kr, and Xe is equivalent to saying that these noble gases have been separated from hydrogen. One way this could be done would be by quantitative condensation onto grains and planetesimals at very low temperatures, probably no higher than 25 K \cp{Owen99}.  Such a scenario would seem to require that much or most of Jupiter's core mass accreted from these very cold objects, otherwise the less volatile N, C, and S would be significantly more abundant than Ar, Kr, and Xe.  Other pathways towards the enhancement of the heavy nobles gases have also been postulated.  Another hypothesis involves bringing the noble gases to Jupiter and Saturn via clathrate hydrates \cp{Gautier01a,Hersant08}.  Alternatively, \ct{Guillot06a} suggest that Jovian abundance ratios are due to the relatively late formation of the giant planets in a partially evaporated disk.  These theories make their own specific predictions for the abundances of the nobles gases. (See Figure 2.)  A completely different possibility is that Jupiter's interior excludes the heavier noble gases, sulfur, nitrogen, and carbon more or less equally, so that in a sense Jupiter would have an outgassed atmosphere\footnote{It is very difficult to diffusively separate H$_2$ and He from the other gases; the effect is weak enough that it would take more than a billion years to diffusively separate a Jupiter's mass of hydrogen from an area comparable to the present Solar System if solar gravity is the force.}.

It is possible to obtain abundances of N, C, and S remotely through optically active molecules such as NH$_3$, CH$_4$, and H$_2$S.  \emph{But the only way to address noble gas abundances in giant planets is by probe.}  A Saturn probe provides the best test of the competing possibilities.  For instance, in the clathrate hydrate hypothesis, \ct{Hersant08} use a solar nebula model to predict in Saturn enhanced Xe, due to its condensation, but a solar abundance for Ar and Kr, which would need lower temperatures to condense.  In the Owen et al.~cold condensate hypothesis, evidence that carbon in Saturn is more than twice as abundant as it is in Jupiter would imply that Ar, Kr, and Xe would also be more than twice as abundant in Saturn.  If the cold condensate hypothesis is correct, it has profound importance for understanding solar nebular evolution and giant planet formation. 

\section{Detecting Noble Gases}
The \emph{remote sensing} measurement of Saturn's atmospheric helium abundance has been fraught with difficulty, elaborated in great detail in \ct{CG00}.  The standard remote sensing method involves obtaining an atmospheric \emph{P-T} profile from radio occultations, along with mid-infrared spectra in regions where H$_2$ collision induced absorption is important.  This is because H$_2$/He collisions help to shape the opacity of the H$_2$.  A radiative transfer model is run, using the empirical \emph{P-T} profile, and the He/H$_2$ ratio of the model is adjusted until the best match to the spectrum is achieved.  Of critical importance is that this method apparently failed for Jupiter, as the \emph{Voyager} remote sensing value of $Y$, the He mass fraction, $Y=0.18 \pm 0.04$ \cp{Gautier81} does not agree with the \emph{Galileo Entry Probe} value of $Y=0.234 \pm 0.005$ \cp{vonzahn98}\footnote{We note that even this value is substantially depleted relative to $Y_{\rm protosolar}$, which is constrained from helioseismology to be $Y=0.270 \pm 0.005$ \cp{Asplund09}.}.  The Saturnian \emph{Voyager} value of $Y$ is extremely low, $Y=0.06 \pm 0.05$ \ct{Conrath84}, but is now thought to be incorrect.  The \ct{CG00} reanalysis of the \emph{Voyager} data sets, with a somewhat different method, put Saturn's atmospheric $Y$ at 0.18-0.25, but the community has only modest confidence in this number.

A dedicated $Y$ measurement on a Saturn entry probe \cp[e.g.][]{vonzahn98} would help to provide a resolution of the uncertainty.  To separate chemical effects in Saturn's interior from possible mass-dependent effects, an additional measurement of isotopic ratios HD/H$_2$ and He$^3$/He$^4$ by means of in situ mass spectroscopy \cp{Niemann98} would be important.  Furthermore, there is clearly no way to remotely measure the abundances of the other noble gases, which are trace components.  \emph{A Saturn entry probe is essential to measuring the abundances of these gases.}

\section{Helium, Jupiter, and Saturn}
A credible, complete understanding of the thermal evolution of Jupiter and Saturn cannot be claimed until the atmospheric helium abundance is known in Saturn.  A measurement of $Y$ by a Saturn entry probe would help to close the door on what is by now a $\sim$35-year-old, basic problem in solar system science.  Now is an excellent time to be taking a new look at the evolution of these planets, as the rise of modern high-pressure shock experiments and supercomputers have finally allowed for the calculation and testing of accurate first-principles H/He equations of state \cp[e.g.][]{Militzer08}.  Previous experimental and theoretical uncertainties have been greatly reduced.  Three-dimensional simulations of convective energy transport in the face of helium (and any other) composition gradients in the deep interiors of giant planets (P.~Garaud, and collaborators, UCSC) should greatly diminish the last remaining uncertainty in giant planet evolution models \cp{SS77a,FH03}.

As mentioned above, the detection of other nobles gases in the atmosphere of Saturn would bear strongly on the formation of the planet.  But another noble gas is relevant to the helium phase separation issue.  That is neon, which like helium is depleted in the atmosphere of Jupiter.  It has been suggested that neon dissolves into the phase-separated helium droplets, and is lost to deeper layers in Jupiter and Saturn \cp{Roulston95}.  However, these calculations need to be confirmed by other physicists, and a measurement of depleted neon in Saturn's atmosphere would be an important data point in understanding this process.

A combination of precise values of the helium (and neon) abundance in Jupiter and Saturn, together with the new EOSs, and advanced cooling models, will allow for an accurate and precise understanding of the helium distribution and temperature distribution within these planets.  A consistent evolutionary history for both of these planets can be obtained.  Models of the current structure, constrained by the planets' gravity fields, will also be impacted.  A better constraint on the distribution of helium within Saturn will lead directly to a better constraints on Saturn's core mass \cp{Hubbard05}, as was previously achieved for Jupiter \cp{Guillot97}.

\section{The Exoplanet Connection}
Inside the solar system 1 of our 2 gas giant planets does not fit within the simple homogeneous picture of planetary cooling.  Outside of the solar system, our ``standard'' theory of giant planet cooling has failed again, this time in explaining the radii of the transiting hot Jupiters.  Over 50 transiting planets have now been published, and \emph{$\sim$40\% of these planets} have radii larger than can be accommodated by standard cooling models \cp{Fortney07a,Baraffe08,Miller09}.  There are numerous explanations for what might be causing this major discrepancy, which we will not detail here.

The direct imaging of giant planets is now taking off as well, as five planets around three parent stars were recently published \cp{Marois08,Kalas08,Lagrange09}.  While transiting planets have an easily measurable mass and radius, the same is not true for these planets.  Here accurate cooling models are even more vital.  Mass estimates come \emph{only from comparison to cooling models}.  These models aim to predict the luminosity, radius, and infrared spectra and colors as a function of mass and age \cp[e.g.][]{Burrows97,Chabrier00b,Marley07,Fortney08b}.  In order to understand the evolution of exoplanets, it is critical to get better ``ground truth'' points to tie on to, namely Jupiter and Saturn, so that we can examine exoplanet properties with greater confidence.

The importance of solar system ground truth surely extends to formation as well.  Incredibly, giant planets in extrasolar systems are found from distances from from 0.015 AU (the closest-in hot Jupiters) to 100 AU (the directly imaged planets).  Detailed data to constrain solar system planet formation will allow us to better understand if all giant planets do generally share a common formation mechanism, and perhaps if all parts of this mechanism are common.

\section{The Fundamental Physics Connection}
Studying the physics of hydrogen and helium is attractive for many reasons.  Perhaps most importantly, they \emph{can} be studied, as these elements are the simplest and most abundant in the Universe.  Techniques, tools, and theories are often honed on these elements before physicists move onto the heavier elements. Also, these elements are important for fusion, and the United States has put many billions of dollars into fusion science over the decades.  Most recently, LLNL is now commissioning the National Ignition Facility, a \$4.2 billion facility to make progress towards H-fusion (and other projects) using pulsed laser power.

For decades, physicists have investigated under what pressure-temperature conditions H and He will phase separate.  All early efforts \cp{SS77b,HDW,Pfaff} had to make a number of simplifying assumptions in order to even attempt to solve the problem.  But with the rise of modern supercomputers, this problem can now be attacked using first-principles methods, in its full quantum mechanical glory.  Importantly, two very recent papers \cp{Morales09,Lorenzen09} provide strong evidence for phase separation of the H/He mixtures in the interior of Saturn, and probably also of Jupiter.  They predict critical temperature for demixing at $\sim$~8000 K, which is certainly reached in Saturn.

These phase diagrams cannot be studied experimentally, as the necessary pressure-temperature regime cannot yet be reached, and there is no clear path towards actually measuring phase separation in laser-induced dynamic shock experiments that last only tiny fractions of a second.  Jupiter and Saturn are our natural laboratories where the effects of phase separation have been playing out for gigayears.

\section{Conclusion}
In this brief paper, we have highlighted the importance of obtaining an accurate and precise measurement of the noble gas abundances in Saturn's atmosphere.  These can only be obtained by a Saturn entry probe.  For helium in Saturn and Jupiter, we will finally achieve a proper understanding of the extent and physical effects of helium phase separation, and a new generation of thermal evolution and structural models will yield a great leap forward in our understanding of these planets.  For the heavier noble gases, their abundances will constrain formation scenarios for these planets.  Jupiter and Saturn serve as the \emph{calibrators} for our understanding of the formation and evolution of all giant planets, who now number over 300 in exoplanetary systems.  Accurate models are fundamental to understanding these planets.  In addition, Saturn's helium abundance, along with Jupiter's, are the only constraints yet possible on the physical interaction of H/He at megabar pressures, an important regime of condensed matter physics.


\end{document}